\documentclass[cits,a4paper]{PoS}
\usepackage{amsmath,amssymb}
\usepackage[tight,TABTOPCAP]{subfigure}
\usepackage{graphicx}
\def\fm{{\rm fm}}
\newcommand{\nft}{N_\mathrm{f}=2}
\newcommand{\nfto}{N_\mathrm{f}=2+1}
\newcommand{\nftoo}{N_\mathrm{f}=2+1+1}
\newcommand{\nfth}{N_\mathrm{f}=3}
\newcommand{\nff}{N_\mathrm{f}=4}
\newcommand{\Oa}{\mathcal{O}(a)}
\newcommand{\Oasq}{\mathcal{O}(a^2)}
\newcommand{\mev}{\mathrm{MeV}}

\newcommand{\mps}{m_\mathrm{PS}}
\newcommand{\fps}{f_\mathrm{PS}}
\newcommand{\mpcac}{m_\mathrm{PCAC}}

\newcommand{\preprint}{\newline%
  \begin{picture}(0,0)
    \put(240,100){\rm\small DESY 11-025, FTUAM-11-37, SFB/CPP-11-09}
  \end{picture}
}

\graphicspath{{plots/}}

\title{Towards Four-Flavour Dynamical Simulations \preprint}

\ShortTitle{Towards Four-Flavour Dynamical Simulations}

\author{\speaker{Gregorio Herdoiza}\\
  NIC, DESY\\
  Platanenallee 6, 15738 Zeuthen, Germany\\
  {\rm and}\\
  Departamento de F\'isica Te\'orica and
  Instituto de F\'isica Te\'orica UAM/CSIC\\
  Universidad Aut\'onoma de Madrid, Cantoblanco 28049 Madrid, Spain\\
  E-mail: \email{Gregorio.Herdoiza@desy.de} }

\abstract{The inclusion of physical effects from sea quarks has been
  one of the main advances in lattice QCD simulations over the last
  few years. We report on recent studies with four flavours of
  dynamical quarks and address some of the potential issues arising in
  this new setup. First results for physical observables in the light,
  strange and charm sectors are presented together with the status of
  dedicated simulations to perform the non-perturbative
  renormalisation in mass-independent schemes.}

\FullConference{The XXVIII International Symposium on Lattice Field
  Theory, Lattice2010\\ June 14-19, 2010\\ Villasimius, Italy}

\begin{document}

\section*{Introduction}

Non-perturbative studies of QCD on the lattice are steadily
approaching a stage in which all systematic uncertainties present in
the numerical calculations can be reliably estimated. A major advance
during last decade has been to surpass the quenched approximation by
the inclusion of sea quark effects.
Dynamical simulations with $\nft$ flavours of mass-degenerate up and
down sea quarks and $\nfto$ calculations where the strange quark is
also incorporated are currently being
performed.
For reviews on the status of these dynamical simulations at this and
other recent lattice conferences, we refer
to~\cite{Hoelbling:2011kk,Jung:2010jt,Scholz:2009yz,Jansen:2008vs}. Investigations
about the difficulties encountered in simulating at small values of
the lattice spacing are reviewed in~\cite{Luscher:2010we}.

The Appelquist-Carazzone theorem~\cite{Appelquist:1974tg} describes
the decoupling of heavy particles at small momentum transfer. Loop
effects from charm quarks should indeed be suppressed with respect to
those of the other lighter sea quarks. At large momentum transfer
charm quarks become active and physical processes could depend in a
non-negligible way on charm loop effects. The renormalisation group
running of renormalised quantities depends on the number of active
flavours. In a mass-independent scheme, a matching between theories
differing by the number of flavours is performed at a scale around the
threshold quark mass above which this quark becomes active.
For energy scales ranging between the charm and the bottom quark
masses it is important to incorporate the effects of the active charm
in the non-perturbative running with $\nff$ massless quarks. When
considering renormalisation group invariant quantities, it is
appropriate to opt for a realistic setup in which the physical
charm-loop effects are included together with those of the lighter
$u$, $d$ and $s$ sea quarks. These $\nftoo$ simulations allow to
explore the effect of dynamical charm on hadron masses and matrix
elements, thus removing this so far uncontrolled systematic effect.

\noindent Incorporating a charm quark in dynamical simulations can however
introduce potential difficulties.
\begin{itemize}
\item The values of the inverse lattice spacing currently in use are
  not considerably larger than the charm quark mass $m_c$.  Lattice
  artefacts proportional to $m_c$ are thus in general significant,
  as it is well known from the study of observables involving charm
  quarks in the valence sector. In $\nftoo$ simulations, cutoff
  effects of this kind, but this time stemming from the fermionic
  determinant, can potentially affect all observables, even those
  without valence charm quarks. An explicit control of the size of
  cutoff effects in quantities that can be measured accurately --
  such as light-quark observables -- in the presence of dynamical
  charm-loops can allow to address this issue.
\item Dynamical simulations aim at setting the sea quark masses as
  close as possible to their physical values through the use of
  appropriate hadronic observables. In the $\nftoo$ case, the amount
  of computing resources dedicated to the tuning of the quark masses
  can be fairly large. It is therefore legitimate to wonder whether
  the tuning effort is prohibitive.
\item Finally, the use of a mass-independent renormalisation scheme
  implies that dedicated simulations with $\nff$ massless, or nearly
  massless, quarks need to be considered. We remark that a similar
  effort is also needed in three-flavour simulations.
\end{itemize}
In the following, we will address these three issues based on recent
four-flavour calculations.
\begin{figure}[t]
  \centering
  \subfigure[\label{fig:a_mpi_211}]{
    \includegraphics[height=0.41\linewidth]{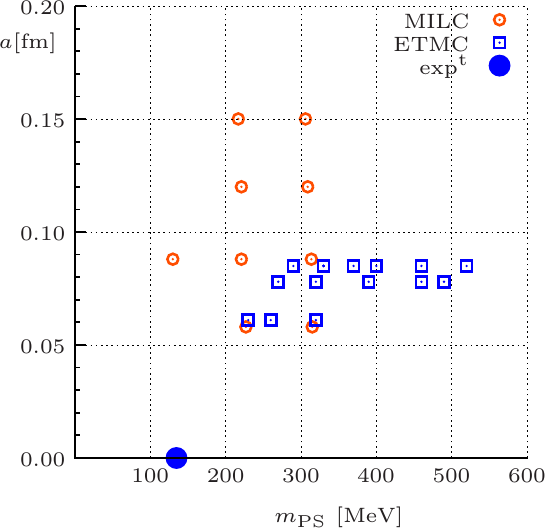}}
  \qquad
  \subfigure[\label{fig:mpi_L_211}]{
    \includegraphics[height=0.41\linewidth]{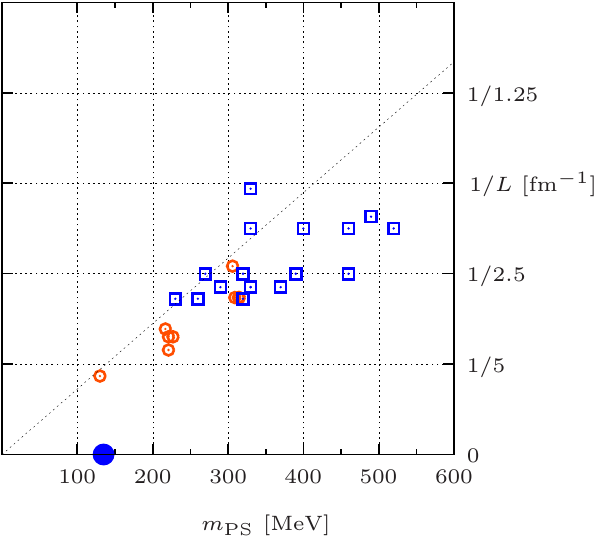}}
  \caption{Parameter space of dynamical simulations with $\nftoo$
    flavours by MILC and ETMC. Simulation points (a) in the lattice
    spacing $a$ and light pseudoscalar mass $\mps$ plane and (b) in
    the inverse lattice size $1/L$ and $\mps$ plane are
    considered. The oblique dotted line refers to the line of constant
    $\mps L=3.5$. The physical point is indicated by the blue circle.}
  \label{fig:landscape}
\end{figure}

\section{$\nftoo$ Simulations~:~Light-Quark Sector}

In this section we review the recent $\nftoo$ simulations by MILC and
ETMC. A particular attention is given to the continuum-limit scaling
of light-quark observables to identify potentially large cutoff
effects due to the heavy mass $m_c$ in the sea sector.

\subsection{MILC Studies with HISQ Quarks}

{\bf \noindent Simulation Setup.} The MILC
collaboration~\cite{Bazavov:2010ru,Bazavov:2010pi} is producing
ensembles with $\nftoo$ flavours of highly improved staggered quarks
(HISQ)~\cite{Follana:2006rc}. In the gauge sector a tadpole-improved
one-loop-Symanzik-improved action is used. The gauge action includes
the effects of the quark loops in the one-loop coefficients. The HISQ
action differs from the asqtad action used in the $\nfto$
simulations~\cite{Bazavov:2009bb} by further smearing of the gauge
links entering in the covariant derivative. HISQ quarks also include
the Naik term, i.e. a third-nearest-neighbour coupling, with a
mass-dependent correction to the tree-level improvement of the
charm-quark dispersion relation. While being computationally more
expensive than asqtad, the HISQ action is expected to reduce taste
symmetry violations through the use of highly smeared links. For
information about the lattice action and the algorithm, such as the
modification of the guiding Hamiltonian to avoid large fluctuations in
the fermion force, we refer to~\cite{Bazavov:2010ru}.

MILC ensembles cover four values of the lattice spacing, $a \approx
\{0.06,\,0.09,\, 0.12,\,0.15\}\,\fm$. The light pseudoscalar meson
mass $\mps$ varies from $315\,\mev$ down to the physical point. The
simulation at the physical pion mass is performed at a lattice spacing
$a \approx 0.09\,\fm$ and is in progress. The lattice size ranges
between $2.4$ and $5.5$~fm and satisfies the condition $\mps L >
3.5$. The red circles in Fig.~\ref{fig:landscape} illustrate the
status of the MILC $\nftoo$ program~\cite{Bazavov:2010pi}. Finalised
ensembles include more than 5000 molecular dynamic units, with
trajectory length ranging between $1.0$ and
$1.5$~\cite{Bazavov:2010ru}.\\

\begin{figure}[t!]
  \vspace*{-1.7cm}
  \centering
  \subfigure[\label{fig:fps_milc}]{
    \includegraphics[width=0.44\textwidth]{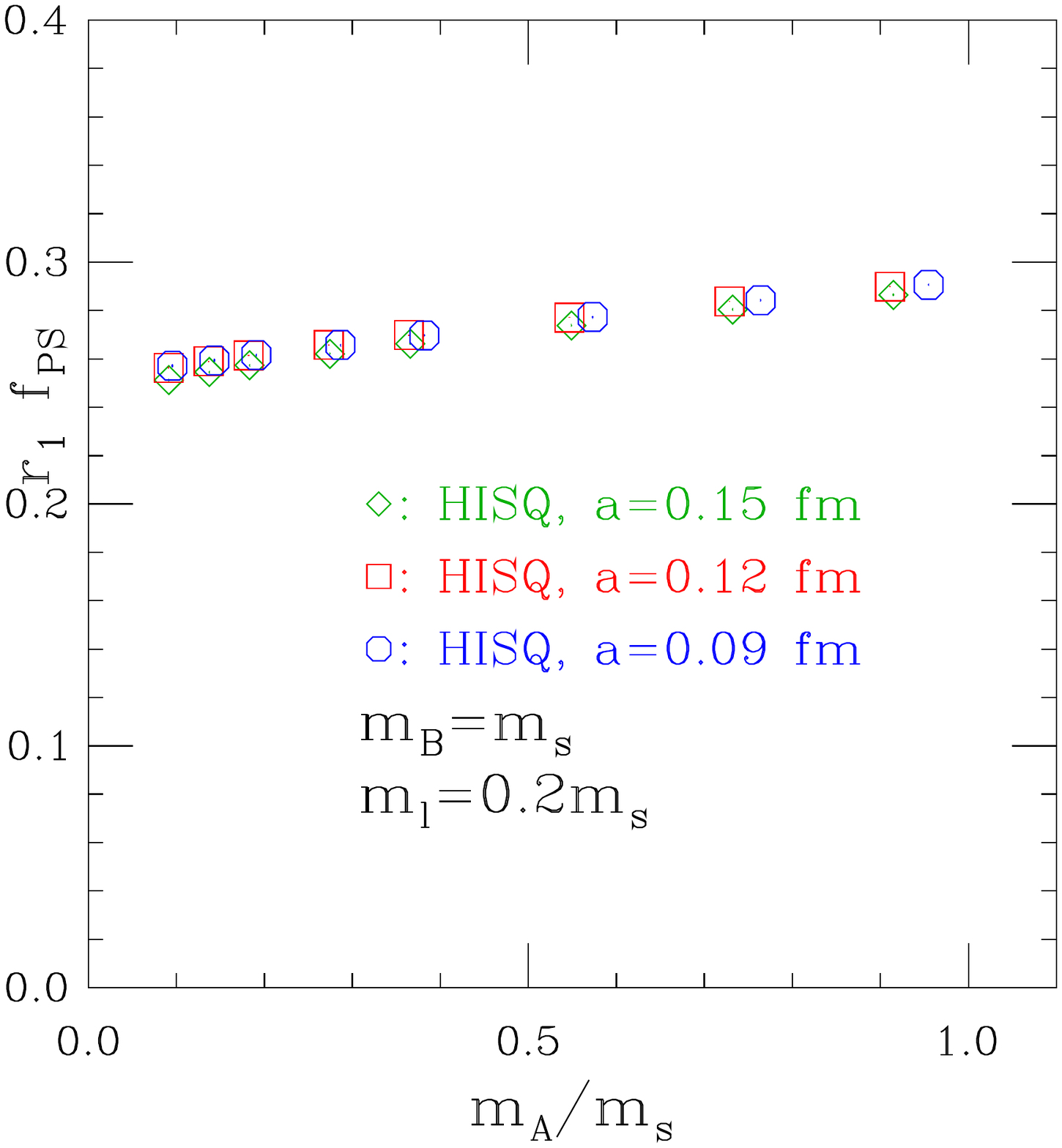}}
  \qquad
  \begin{minipage}[t]{0.40\textwidth} 
    \vspace*{-8.3cm}
    \subfigure[\label{fig:taste_milc}]{
      \includegraphics[width=1.0\textwidth]{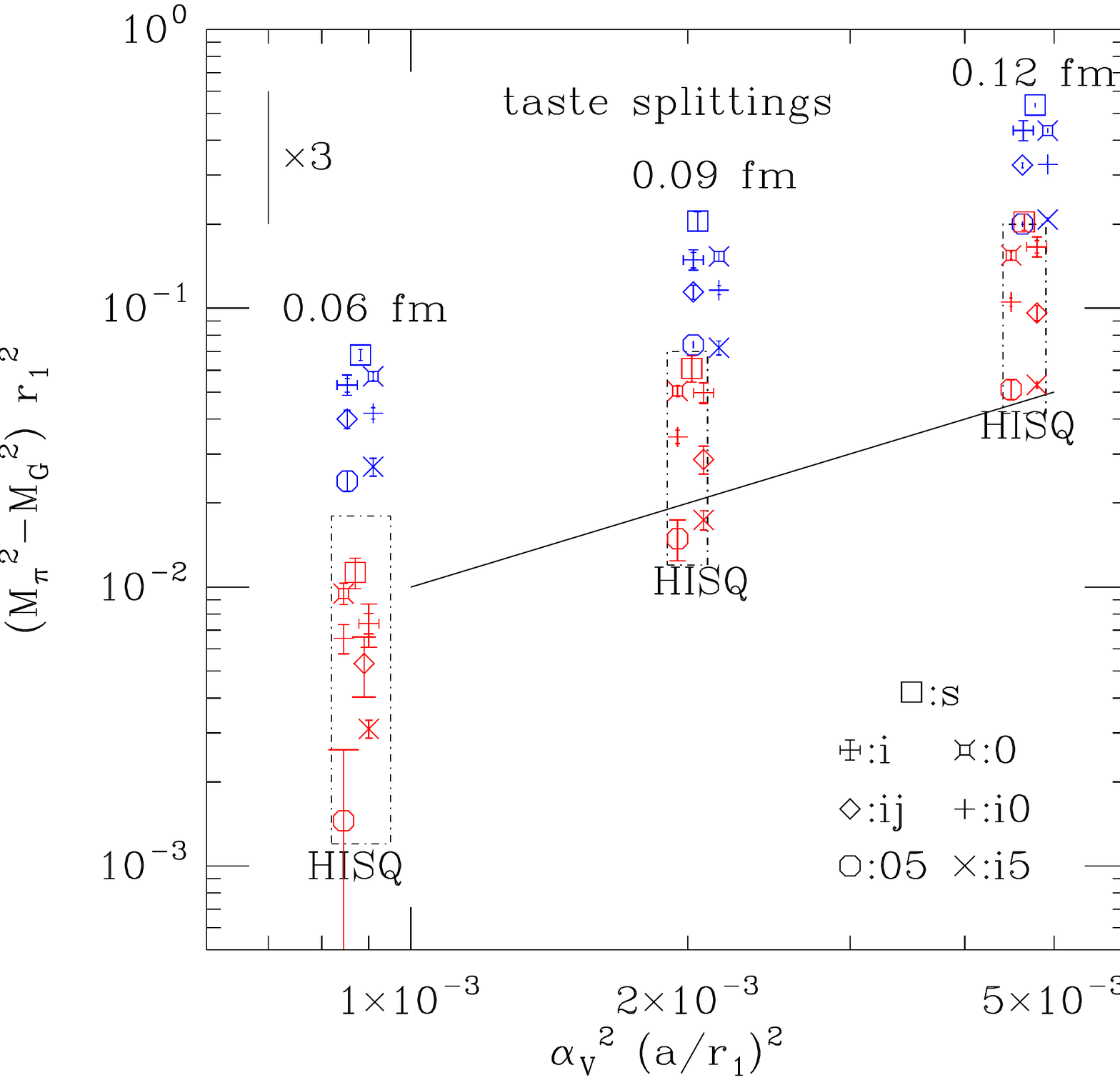}}
  \end{minipage}
  \caption{Continuum-limit scaling studies by MILC. (a) Pseudoscalar
    meson decay constant $\fps$ in units of the scale $r_1$ as a
    function of the light-quark mass $m_A$ (normalised by the strange
    quark mass $m_s$). The pseudoscalar meson is composed of the
    valence quark masses $m_A$ and $m_B$, the latter being fixed to
    $m_B=m_s$. The unitary point corresponds to $m_A/m_s= 0.2$ in the
    x-axis. Small lattice artifacts are observed in the considered
    range of lattice spacings. (b) Continuum-limit scaling of the
    taste splittings of pions. $M_{\rm G}$ stands the mass of the
    lightest (Goldstone) pion, while $M_\pi$ refers to the remaining
    pion masses, organised by taste structure. A reduction of taste
    symmetry breaking is observed when moving from the $\nfto$ asqtad
    action (blue symbols) to the $\nftoo$ HISQ one (red symbols).}
  \label{fig:milc}
\end{figure}

{\bf \noindent Continuum-Limit Scaling.} As shown in
Fig.~\ref{fig:landscape}, MILC has generated ensembles with fixed
pseudoscalar mass $\mps$ and lattice size $L$ at four values of the
lattice spacing. Such a setup is well suited for a continuum-limit
scaling~\cite{Bazavov:2010ru}. The values of the bare charm quark mass
in lattice units vary from $am_c \approx 0.44$ at $a \approx
0.09\,\fm$ up to $am_c \approx 0.84$ at the coarser lattice spacing $a
\approx 0.15\,\fm$~\cite{Bazavov:2010ru,Bazavov:2010pi}. It is
instructive to examine if values of $am_c \lesssim 1$ induce large
lattice artefacts in light-quark observables. The case of the
pseudoscalar meson decay constant $\fps$ is shown in
Fig.~\ref{fig:fps_milc}.
The scale $r_1$~\cite{Bernard:2000gd} is used to relate different
lattice spacings. It is determined in a similar way to the Sommer
scale $r_0$~\cite{Sommer:1993ce}, from the force between static
quarks.  Fig.~\ref{fig:fps_milc} shows no evidences for large cut-off
effects in $r_1 \fps$. Furthermore, compared to $\nfto$ asqtad data, a
reduction of lattice artefacts is observed for $\nftoo$ HISQ
ensembles~\cite{Bazavov:2010ru}. Similar effects are observed for
other quantities such as the vector meson mass, the nucleon mass or
the topological susceptibility~\cite{Bazavov:2010ru,Bazavov:2010pi}.
\\

{\bf \noindent Taste Symmetry Breaking.} Each staggered fermion field
generates four degrees of freedom, the so called tastes. In a theory
with $\nff$ mass-degenerate quarks, these four tastes can be
interpreted as quark flavours. In the $\nftoo$ case, where the sea
quark masses satisfy, $m_u=m_d < m_s < m_c$, a ``rooting procedure''
is applied to the fermionic determinant in order to eliminate the
contribution of unwanted tastes. As a result, locality and unitarity
of the theory do not hold at non-zero lattice
spacing~\cite{Bazavov:2009bb}. Assessing the validity of the approach
once the continuum-limit is taken has triggered a number of studies
(see~\cite{Bazavov:2009bb} and references therein for an account on
the rooting issue). With staggered quarks, mesons appear in 16 species
differing by their taste content. At finite lattice spacing, the
breaking of taste symmetry introduces mass-splittings among these
mesons. When taking the continuum limit, mass-splittings among
non-singlet mesons are expected to vanish. These splittings are
largest for the pion system and typically amount for bigger
discretisation effects than other more standard (i.e. not related to
taste symmetry breaking) lattice artifacts. Determining the size of
taste breaking effects is therefore an important check of the
setup. Fig.~\ref{fig:taste_milc} shows the lattice spacing dependence
of the pion taste splittings.
At fixed lattice spacing a significant reduction (of around a factor
of three) in the size of the taste splittings is observed when
switching from asqtad to HISQ. Since the generation of $\nftoo$
ensembles is in progress~\cite{Bazavov:2010pi}, first physical results
from MILC are expected to appear in the near future.

\subsection{ETMC Studies with Wilson twisted mass (Wtm) Fermions}

{\bf \noindent Simulation Setup.} The ETM collaboration is performing
simulations with $\nftoo$ flavours of Wilson twisted mass (Wtm)
fermions~\cite{Baron:2010bv}.\,\footnote{This setup was first explored
  in~\cite{Chiarappa:2006ae}.} The Iwasaki
action~\cite{Iwasaki:1985we} is used in the gauge sector. This choice
is driven by considerations on the phase structure of Wilson
fermions. With respect to the tree-level Symanzik-improved gauge
action~\cite{Weisz:1982zw}, which was used in $\nft$
simulations~\cite{Boucaud:2007uk,Boucaud:2008xu}, ETMC observes that
in the $\nftoo$ case, the Iwasaki action provides a smoother variation
of observables around the critical
mass~\cite{Baron:2010bv,Farchioni:2004fs}. The fermionic action is
composed of two doublets of Wtm fermions. Note that the use of an even
number of flavours, e.g. $\nftoo$, is a natural choice for Wtm
fermions. The action for the mass-degenerate light doublet ($u$, $d$)
reads~\cite{Frezzotti:2000nk,Frezzotti:2003ni}
\begin{equation}
  \label{eq:sl}
  S_\ell\ =\ a^4\sum_x\left\{ \bar\chi_\ell(x)\left[ D_{\rm W}[U] + m_{0,l}
    + i\mu_\ell\gamma_5\tau_3\right]\chi_\ell(x)\right\}\, ,
\end{equation}
where $\chi_\ell=(\chi_u, \chi_d)$, $D_W$ is the massless Wilson-Dirac
operator, $m_{0,l}$ is the untwisted bare quark mass, $\mu_\ell$ is
the bare twisted light quark mass and $\tau_3$ is the Pauli matrix
acting in flavour space.  For the heavier non-degenerate ($c$, $s$)
doublet the action takes the following
form~\cite{Frezzotti:2004wz,Frezzotti:2003xj}
\begin{equation}
  \label{eq:sf}
  S_h\ =\ a^4\sum_x\left\{ \bar\chi_h(x)\left[ D_{\rm W}[U] + m_{0,h} +
    i\mu_\sigma\gamma_5\tau_1 + \mu_\delta \tau_3
    \right]\chi_h(x)\right\}\, ,
\end{equation}
where $\chi_h = (\chi_c, \chi_s)$, $m_{0,h}$ is the untwisted quark
mass, $\mu_\sigma$ the twisted mass -- the twist is along the
$\tau_{1}$ direction -- and $\mu_\delta$ the mass splitting along the
$\tau_{3}$ direction. The bare mass parameters $\mu_\sigma$ and
$\mu_\delta$ are related to the renormalised strange and charm quark
masses~\cite{Frezzotti:2004wz}:
\begin{equation}
  \label{eq:msmc}
  (m_{s})_{\rm R} = Z_{\rm P}^{-1} \, (\mu_\sigma - Z_{\rm
    P}/Z_{\rm S} \, \mu_\delta)\,, \qquad (m_{c})_{\rm R} = Z_{\rm
    P}^{-1} \, (\mu_\sigma + Z_{\rm P}/Z_{\rm S} \, \mu_\delta)\, ,
  \nonumber
\end{equation}
where $Z_{\rm P}$ and $Z_{\rm S}$ are the renormalisation constants of
the pseudoscalar and scalar non-singlet quark densities, respectively,
computed in the massless $\nff$ standard Wilson theory. The $\Oa$
improvement of physical observables is obtained by working at maximal
twist~\cite{Frezzotti:2003ni,Frezzotti:2004wz}. This is achieved by
imposing~\cite{Chiarappa:2006ae} $am_{0,l}=am_{0,h}\equiv
{1}/{2\kappa} -4$ and by setting $\kappa=\kappa_{\rm crit}$ through
the tuning of the light PCAC quark mass to zero at each set of values
$\{\mu_\ell,\,\mu_\sigma,\,\mu_\delta\}$. For more information about
the lattice setup, the tuning to maximal twist and the algorithm we
refer to~\cite{Baron:2010bv}.

ETMC ensembles include three values of the lattice spacing , $a
\approx \{0.06,\,0.08,\,0.09\}\,\fm$ and lattice extents ranging from
$1.9$ to $2.7$~fm. The light pseudoscalar mass $\mps$ varies from
$520\,\mev$ down to $230\,\mev$. For each point, the largest lattice
size satisfies $\mps L \gtrsim 3.5$. The blue squares in
Fig.~\ref{fig:landscape} illustrate the status of the ETMC $\nftoo$
simulations~\cite{Baron:2011sf}. Ensembles contain 5000 thermalised
trajectories of length $\tau=1$.\\

\begin{figure}[t!]
  \centering
  \subfigure[\label{fig:fps_etmc}]{
    \includegraphics[width=0.42\linewidth]{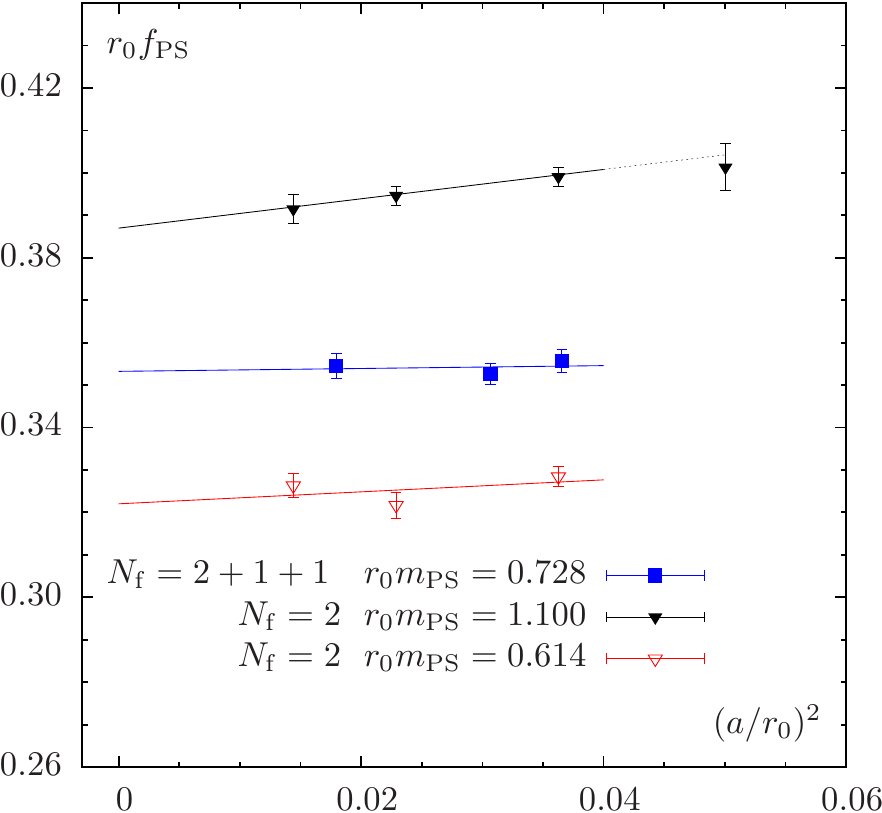}}
  \qquad
  \begin{minipage}[t]{0.46\textwidth} 
    \vspace*{-6.1cm}
    \subfigure[\label{fig:fps_chpt_etmc}]{
      \includegraphics[width=1.0\linewidth]{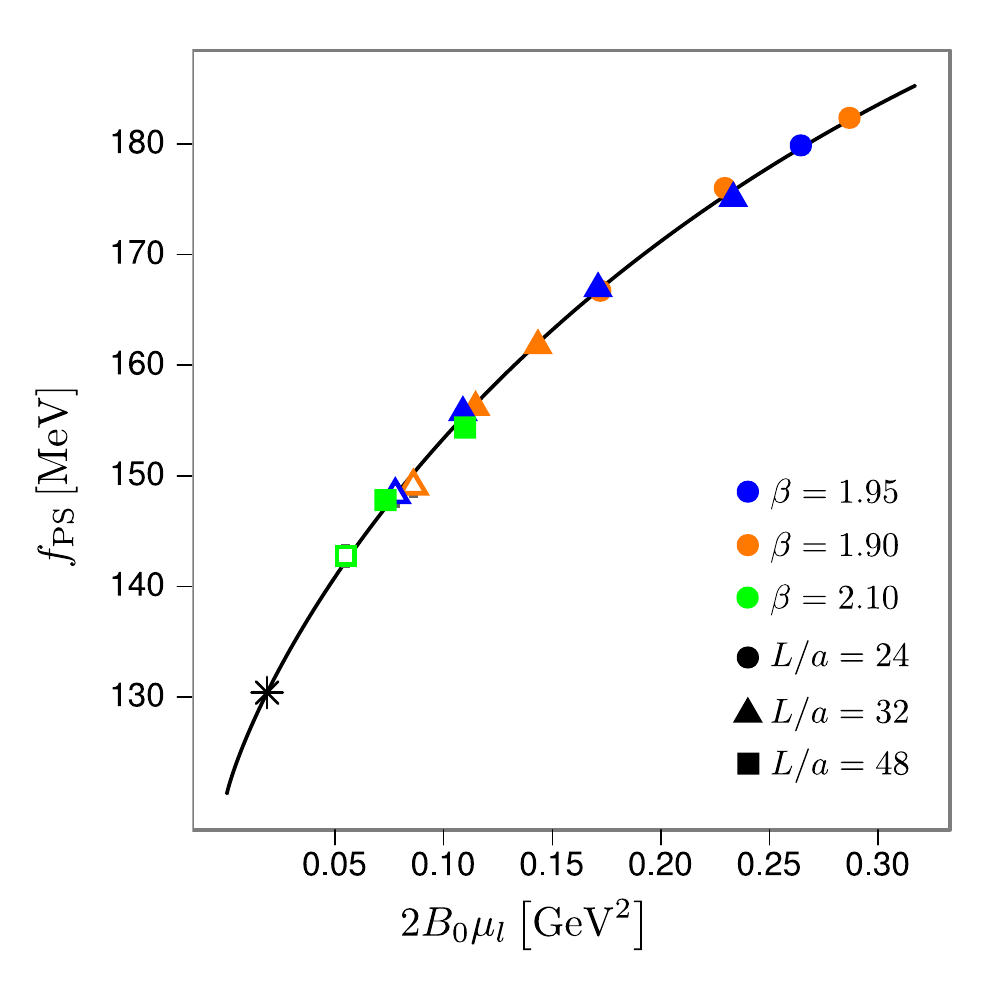}}
  \end{minipage}
  \caption{(a) ETMC study of the continuum-limit scaling of the light
    pseudoscalar meson decay constant $\fps$ in units of the Sommer
    scale $r_0$, at fixed reference values of $\mps$. Mild cutoff
    effects are observed and a comparison to $\nft$ ETMC data
    indicates a similar scaling behaviour for $r_0 \fps$. (b) $\fps$
    as a function of the light quark mass ($B_0$ is the LO $\chi$PT
    LEC). A continuum NLO SU(2) $\chi$PT fit to data from three values
    of the lattice spacing, $a \approx \{0.06,~0.08,~0.09\}\,\fm$, was
    used to estimate systematic effects in the determination of the
    LECs~\cite{Baron:2011sf}. The experimental value pion decay
    constant (black star) was used to set the scale.}
  \label{fig:scaling_etmc}
\end{figure}

{\bf \noindent Continuum-Limit Scaling.} The lattice spacing
dependence of the charged pion decay constant $\fps$ is illustrated in
Fig.~\ref{fig:fps_etmc}. The scaling is consistent with the expected
$\Oa$ improvement. A comparison to the $\nft$ case~\cite{Baron:2009wt}
is also shown. A good scaling behaviour with no signs of large cutoff
effects due to the dynamical charm is observed in $\fps$, as well as
in the nucleon mass~\cite{Drach:2010hy}. The values of the bare charm
quark mass in lattice units satisfy $am_c \lesssim 0.3$.  As
previously discussed, the physical effect from charm-loops should be
small and suppressed with respect to that of the other lighter
quarks. It is therefore expected that cutoff effects from dynamical
charm-quarks appear as a correction to this small effect.
\\

{\bf \noindent Physical Results in the Light Sector.} The light-quark
mass and the volume dependence of $\mps$ and $\fps$ have been studied
by means of chiral perturbation theory ($\chi$PT). At next-to-leading
order (NLO) in continuum SU(2) $\chi$PT~\cite{Gasser:1983kx}, four
low-energy constants (LEC), $B_0$, $f_0$, $\bar{l}_{3,4}$, contribute
to the mass dependence of $\mps$ and $\fps$. Finite size effects (FSE)
were corrected by the resummed L\"uscher formulae combined with
$\chi$PT~\cite{Colangelo:2005gd}. The central values and statistical
errors reported by ETMC~\cite{Baron:2010bv,Baron:2011sf} have been
determined from a fit to a single lattice spacing, $a \approx
0.08\,\fm$. Systematic effects can arise through lattice artifacts,
FSE or higher order terms in $\chi$PT. Fig.~\ref{fig:fps_chpt_etmc}
illustrates a $\chi$PT fit to $\fps$ for three values of the lattice
spacing. Although a full account of systematic uncertainties is still
missing, we refer to~\cite{Baron:2010bv,Baron:2011sf} for a report on
the current estimates.  The resulting $\nftoo$ determinations from
$\chi$PT fits are $f_\pi/f_0=1.076(3)$, $\bar{l}_{3}=3.70(27)$,
$\bar{l}_{4}=4.67(10)$~\cite{Baron:2010bv,Baron:2011sf}. A good
agreement is observed when comparing these determinations to $\nft$
ETMC results~\cite{Baron:2009wt} and to other lattice
calculations~\cite{Colangelo:2010et}.  Studies of other light-quark
observables, such as the nucleon and $\Delta$
masses~\cite{Drach:2010hy} or nucleon matrix
elements~\cite{Dinter:2011jt} were presented at this conference.\\

\begin{figure}[t!]
  \centering
  \subfigure[\label{fig:mpi0_etmc}]{
    \includegraphics[width=0.45\linewidth]{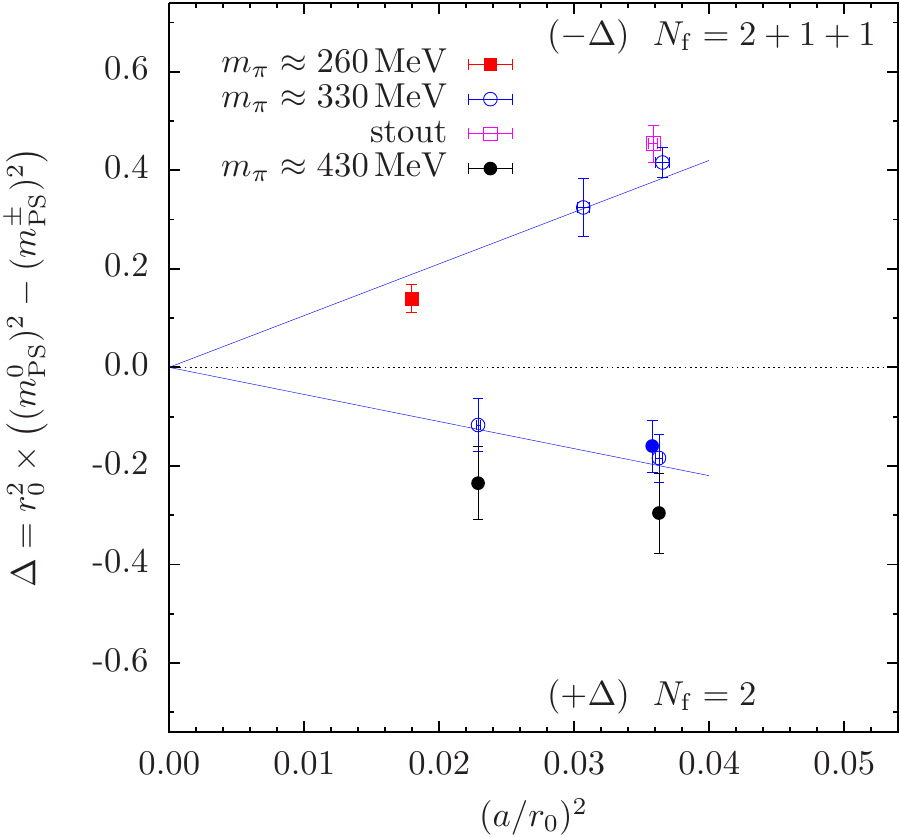}}
  \qquad
  \subfigure[\label{fig:eigD_etmc}]{
    \includegraphics[width=0.45\linewidth]{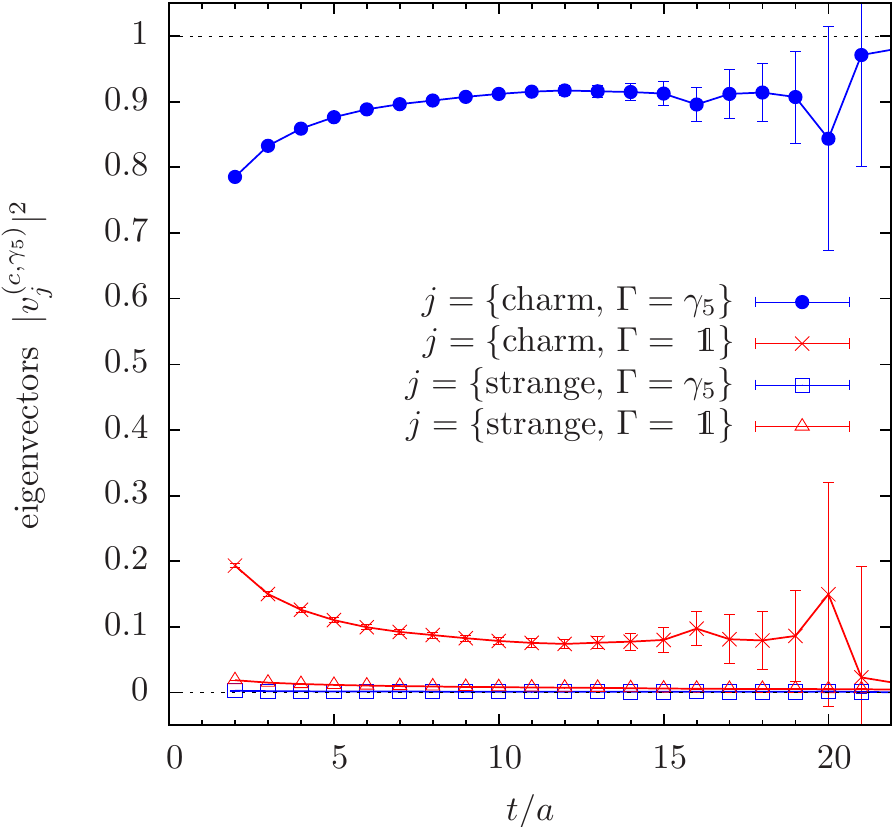}}
  \caption{
    (a) Continuum-limit scaling of the pion
    mass splitting $\Delta$ in units of $r_0$ for $\nftoo$ and $\nft$
    flavours of twisted mass fermions. For a better visibility a
    global minus sign is given to the $\nftoo$ data. The blue lines
    suggest the expected behaviour towards the continuum limit. An
    increase of isospin breaking effects are observed when increasing
    the number of dynamical flavours. The empty square refers to a
    $\nftoo$ ensemble with one level of stout smearing presenting a
    similar mass splitting than a `non-stout' ensemble.
    (b) Approximate flavour and parity content of an state
    extracted from solving the GEP at $a \approx 0.06\,\fm$. The state
    is dominated by the $D$-meson quantum numbers.}
  \label{fig:chi_mpi0_etmc}
\end{figure}

{\bf \noindent Isospin Breaking.} The Wtm action breaks isospin at
finite lattice spacing inducing a mass splitting between charged and
neutral pions. This splitting is a lattice artifact which is expected
to vanish in the continuum at a rate of $\Oasq$. In the $\nft$ case,
ETMC observed mild discretisation effects in the continuum-limit
scaling of the charged pseudoscalar meson mass
$\mps^\pm$~\cite{Baron:2009wt,Blossier:2010cr} while significant
cutoff effects were present in the mass splitting $\mps^0 - \mps^\pm
$~\cite{Baron:2009wt}. Recent studies~\cite{Baron:2010bv}, indicate
that this splitting increases when increasing $N_{\rm f}$ from two to
four.\,\footnote{Note however that the gauge action changed between
  the $\nft$ to $\nftoo$ calculations.}
In the expression $r_0^2((\mps^0)^2 - (\mps^\pm)^2 ) = c\, (a/r_0)^2$,
the sign of the factor $c$ is related to the type of scenario expected
for the phase structure of Wilson fermions in the regime of small
quark masses~\cite{Aoki:1983qi,Sharpe:1998xm}. When $c$ is negative,
it measures the strength of the first order phase
transition~\cite{Munster:2003ba,Scorzato:2004da,Sharpe:2004ny}.
The determination of the $\mps^0$ involves quark-disconnected
contributions. Fig.~\ref{fig:mpi0_etmc} shows the scaling of the pion
mass splitting for $\nft$ and $\nftoo$ ETMC simulations.
The values of $c$ are $c \approx -6$ in the $\nft$ case and $c \approx
-10$ in the $\nftoo$ one.
Besides the pion mass, isospin breaking cutoff effects can effect
other observables. For $\nft$ ensembles, large isospin breaking
effects have been observed only in the neutral pion
mass~\cite{Baron:2009wt}, in agreement with an analysis based on the
Symanzik expansion~\cite{Frezzotti:2007qv,Dimopoulos:2009qv}.
A similar scenario for isospin breaking cutoff effects is expected to
hold in the $\nftoo$ case. Explicit control of isospin breaking, in
particular by performing the continuum extrapolation, is the
appropriate way of addressing this issue. First analyses of the
$\Delta$ baryon masses indicate that isospin breaking effects are
negligible at the three values of the lattice
spacing~\cite{Drach:2010hy}.

The breaking of isospin and parity can be described in Wilson twisted
mass chiral perturbation theory
(tmW$\chi$PT)~\cite{Munster:2003ba,Scorzato:2004da,Sharpe:2004ny}. The
impact of $\mps^0$ in the volume~\cite{Colangelo:2010cu} and
quark-mass dependence~\cite{Bar:2010jk} of the charged pion mass and
decay constant has recently been studied in tmW$\chi$PT. The quality
of the fits is reported to improve~\cite{Colangelo:2010cu,Bar:2010jk}
when including isospin breaking effects. For $\nft$ data, the impact
of these corrections on the determination of the light-quark mass was
observed to be at most at the level of the fitting
errors~\cite{Blossier:2010cr}. ETMC has planned a dedicated study on
this issue in the~$\nftoo$ case.

\section{Strange and Charm Sectors}

{\bf \noindent Tuning Effort.} The simulated quark masses are to be
set as close as possible to their physical values. For the light ($u$,
$d$) quarks this usually implies a chiral extrapolation though recent
progress in the field goes precisely in the direction of simulating
directly at the physical point~\cite{Hoelbling:2011kk}. The strange
and charm quark masses are directly accessible on current
lattices. Since a reliable estimate of the lattice spacing is only
available a posteriori, setting $m_s$ and $m_c$ typically requires
considering a set of values such that a small interpolation to the
physical point can be performed. Such a procedure is also required to
address the non-negligible systematic effect associated to different
ways of setting the scale on the lattice. The inclusion of a dynamical
charm implies a further computational effort with respect to $\nfto$
simulations, due to the additional tuning of $m_c$. When aiming at an
interpolation to the charm quark mass, an increase by roughly a factor
of two in computing time is needed with respect to $\nfto$
calculations. The tuning effort depends on the sensitivity of the
observables of interest to the quark mass and on the target accuracy
one aims to achieve. Reweighting techniques can be useful to perform
the small quark mass corrections in the strange and charm sectors~(see
\cite{Jung:2010jt} for a review). As discussed in
Section~\ref{sec:npr}, the non-perturbative renormalisation requires
dedicated simulations with $\nff$ mass-degenerate flavours. Hence,
carrying out an $\nftoo$ simulation program implies an overall
increase in the computational effort with respect to the $\nft$ and
$\nfto$ cases.\,\footnote{As mentioned in Section~\ref{sec:npr}, we
  note that in the $\nfto$ case, dedicated simulations for
  non-perturbative renormalisation with $\nfth$ flavours should also
  be included in the overall computational cost.} Recent studies with
four dynamical flavours also indicate that this additional effort is
not a fundamental limitation.

To set $m_s$ and $m_c$, MILC uses the physical values of $2m_K^2 -
m_\pi^2$ and $(m_{\eta_c} + 3m_{J/\Psi})/4$,
respectively. Quark-disconnected contributions are not included in the
determination of the charmonium masses but recent
studies~\cite{Levkova:2010ft} indicate that they should amount to a
sub-percent effect on the spin-averaged mass. In the context of
$\nfto$ simulations, QCDSF-UQKCD~\cite{Bietenholz:2010si} is using an
alternative method to set the strange quark mass. Rather than keeping
the strange quark mass fixed when performing the light-quark mass
extrapolation, it is the sum of the quark masses $m_u+m_d+m_s$ that is
kept fixed. A mild dependence of singlet quantities is then observed
when extrapolating data from the SU(3) symmetric point towards the
physical point.\\

{\bf \noindent ETMC Studies of Kaon and $D$-meson Masses.} The twisted
mass action (\ref{eq:sf}) for the ($c$,$s$) doublet is non-diagonal in
flavour space.\,\footnote{This is most easily seen when rewriting the
  action in eq.~(\ref{eq:sf}) in the physical quark basis.} This
induces, at finite lattice spacing, a mixing between strange and charm
quarks. The breaking of flavour symmetry and parity imply that the
identification of the $D$-meson state is not straightforward since its
signal appears among other excited states at intermediate Euclidean
time separations. The ground state is dominated by the Kaon and its
identification is therefore unambiguous. Isolating the $D$-meson by
resolving all the excited states above the Kaon would require large
correlation matrices with high statistical precision. However, in the
$D$-meson channel, the coupling to low-lying states allowed by mixing is
a pure cut-off effect and should therefore be progressively suppressed
towards the continuum limit. At finite lattice spacing, the $D$-meson
state is expected to have a significant contribution to the signal in
the appropriate correlation functions at intermediate time
separations.

Based on this observation, ETMC has developed three methods to
identify the $D$-meson state and to determine its
mass~\cite{Baron:2010th,Baron:2010vp}. These methods rely on (i)
solving a generalised eigenvalue problem (GEP), (ii) fitting the
correlation matrix by a series of exponentials and (iii) enforcing
flavour and parity restoration at finite lattice spacing. We refer
to~\cite{Baron:2010th} for a detailed description of the different
methods. Fig.~\ref{fig:eigD_etmc} shows the approximate heavy flavour
and parity contents of an excited state as determined from the GEP,
i.e. method (i), suggesting that it is dominated by the quantum
numbers of the $D$-meson state. Consistent results are found for the
$D$-meson mass determined by means of the three
methods~\cite{Baron:2010th,Baron:2010vp}. The restoration of parity
and flavour symmetry can be monitored through the continuum-limit
scaling of the elements of the correlation matrix from which the
$D$-meson mass is extracted. A scaling study of the $D$-meson mass in the
free theory has recently been reported in~\cite{Luschevskaya:2010hf}.

\begin{figure}[t!]
  \centering
  \subfigure[\label{fig:mk_etmc}]{
    \includegraphics[height=0.41\linewidth]{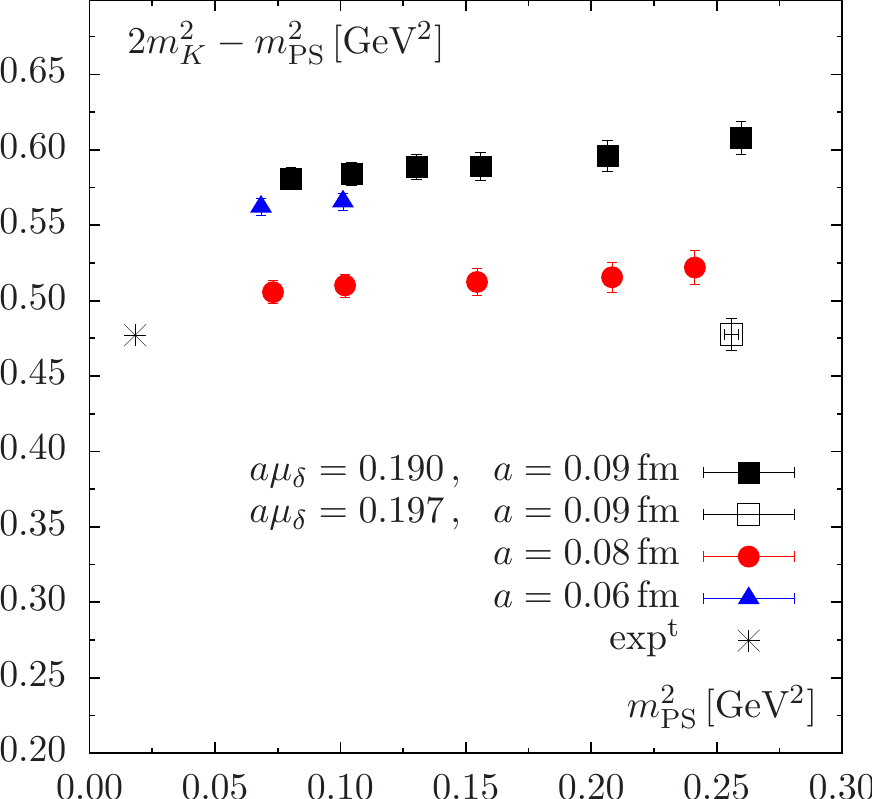}}
  \qquad
  \subfigure[\label{fig:mD_etmc}]{
    \includegraphics[height=0.41\linewidth]{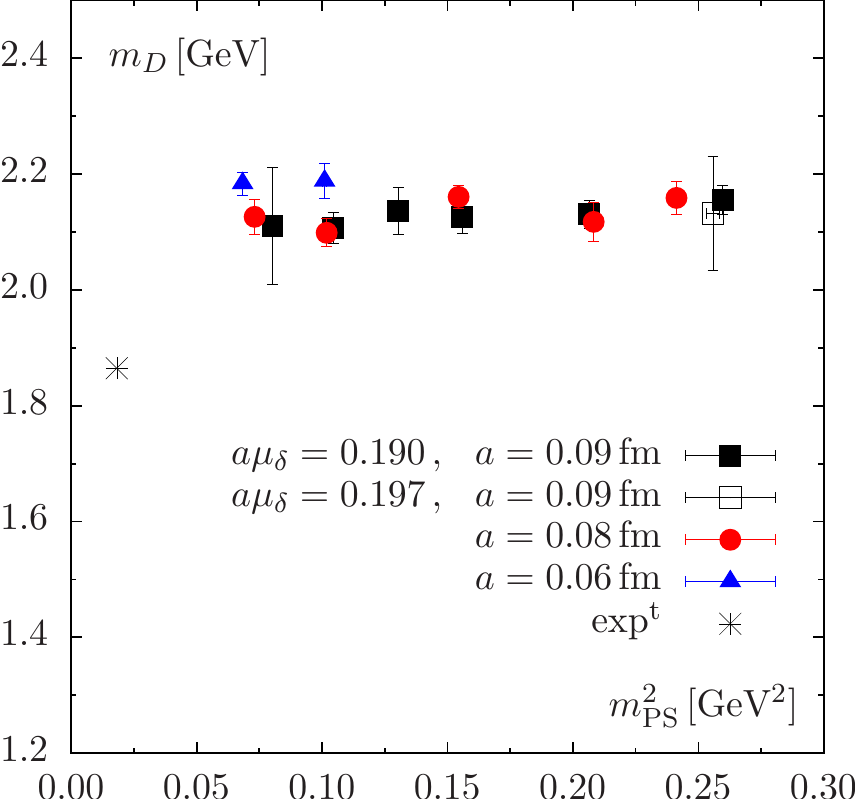}}
  \caption{Status of the tuning of (a) the strange and (b) the charm
    quark mass in $\nftoo$ ETMC ensembles with $a \approx
    \{0.06,\,0.08,\,0.09\}\,\fm$. The physical point is denoted by the
    black star.}
  \label{fig:ms_mc_etmc}
\end{figure}

The status of the tuning of $m_s$ and $m_c$ in ETMC ensembles, through
the use of $2m_K^2-m_\pi^2$ and $m_D$, respectively, is shown in
Figs.~\ref{fig:ms_mc_etmc}. The simulated points are in the
neighbourhood of the physical points (black stars) with maximal
deviations from the physical values of $m_{s,c}$ of the order of $\sim
20\%$. Further tuning runs aiming to interpolate $m_{s,c}$ to the
physical point are currently being performed~\cite{Baron:2011sf}.
\\

{\bf \noindent ETMC Studies of a Mixed Action with Osterwalder-Seiler
  (OS) Quarks.}  Precision studies of charm-quark observables with the
action~(\ref{eq:sf}) is complicated by the presence of the $c$-$s$
mixing discussed above. This problem is avoided in the valence sector
by considering a mixed action setup\,\footnote{A mixed action consists
  in employing different discretisations of the Dirac operator for sea
  and valence quarks. It is thus possible to restore in the valence
  sector a symmetry which is broken in the sea sector.} where the
valence $c$ and $s$ quarks are introduced with a flavour diagonal
action. Osterwalder-Seiler (OS) valence
quarks~\cite{Osterwalder:1977pc} can be viewed as the building blocks
of Wtm fermions at maximal twist. The OS action for an individual
quark flavour $q_f$ reads
\begin{equation}
  \label{eq:os}
  S_\mathrm{OS}\ =\  \bar q_f(x)\left[ \gamma_\mu
    \tilde{\nabla}_\mu -i\,  r_f\, \gamma_5  \left( - \frac{a}{2}
    \nabla_\mu^\ast \nabla_\mu + m_{\rm cr} \right) +
    \mu_f \right] q_f(x) \,, \nonumber
\end{equation}
where $r_f$ (here $|r_f|=1$) is the Wilson parameter and $m_{\rm cr}$
the critical mass. By combining two flavours of OS quarks with
opposite signs of $r_f$, e.g. $r_2 = -r_1$, the action of a doublet of
maximally twisted mass fermions of mass $\mu_f=\mu_1=\mu_2$ can be
recovered. Note that the constraint $\mu_1=\mu_2$ is not necessary for
valence quarks whereas it is crucial in the sea sector in order to
guarantee the reality of the fermionic determinant. In fact, the
determinant of the one-flavour OS lattice Dirac operator is in general
complex. The benefits of the OS action are that $\Oa$ improved
physical observables~\cite{Frezzotti:2004wz} can be achieved by using
the same estimates of $m_{\rm cr}$ than in the Wtm case thus avoiding
further tuning effort. OS and Wtm fermions coincide with Wilson
fermions in the massless limit and therefore share the same
renormalisation factors. This simplifies the matching of sea and
valence quark masses in the context of a mixed action. Finally, the OS
action being flavour diagonal, it is a natural choice for avoiding the
previously discussed strange and charm mixing in the valence sector.

\begin{figure}[t!]
  \centering
  \subfigure[\label{fig:mk_nf2_etmc}]{
    \includegraphics[width=0.45\linewidth]{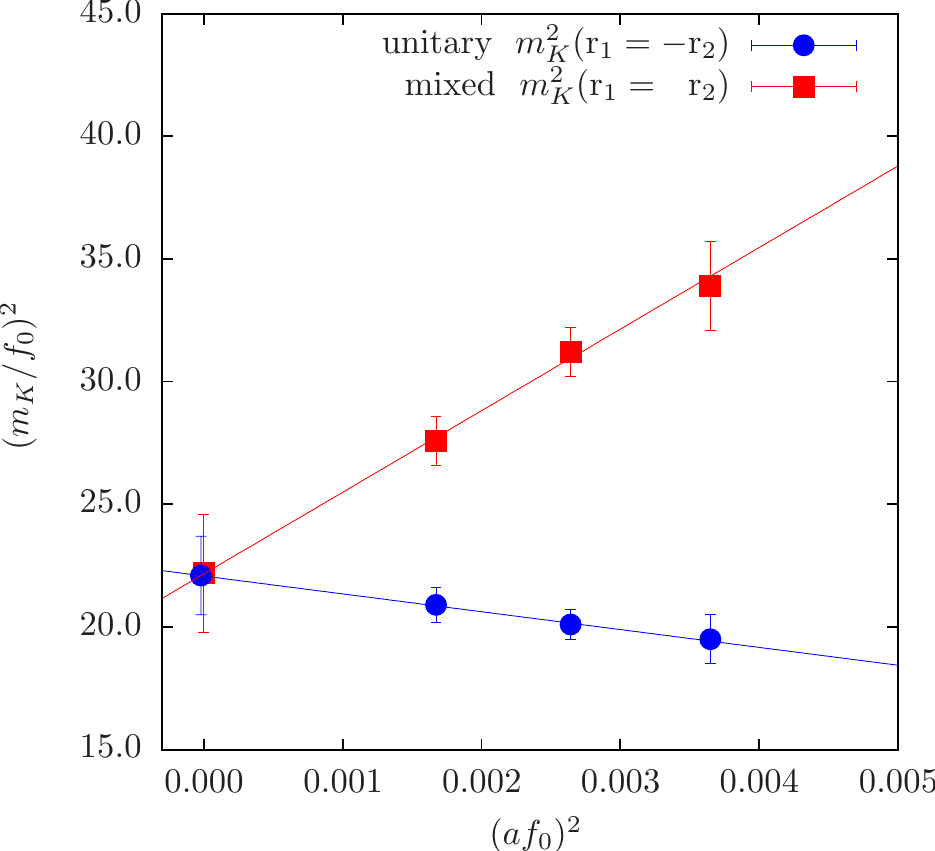}}
  \qquad
  \subfigure[\label{fig:fk_ma_etmc}]{
    \includegraphics[width=0.45\linewidth]{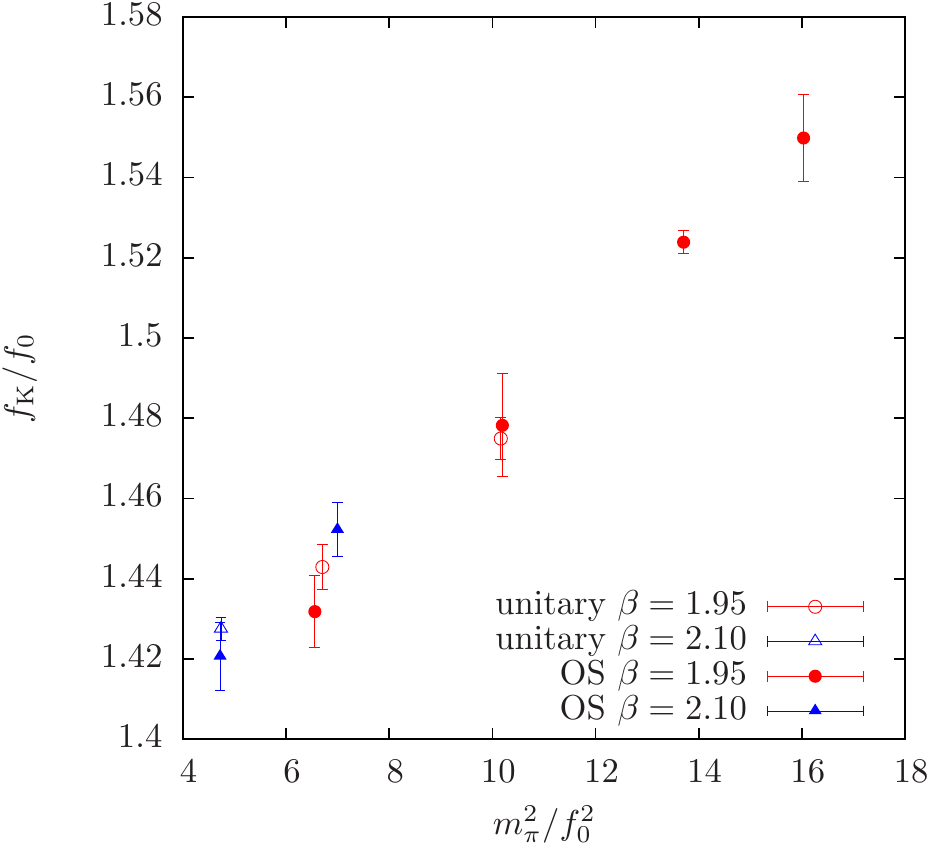}}
  \caption{ETMC studies of mixed actions with OS valence quarks.  (a)
    Continuum-limit scaling of $m_K^2$ for unitary and mixed action
    Kaons in $\nft$ ensembles. The chirally extrapolated pseudoscalar
    decay constant $f_0$ is used as scaling variable. The restoration
    unitarity is observed in the continuum limit. (b) Unitary and
    mixed action $f_K$ from $\nftoo$ ensembles with $a \approx 0.06$
    and $0.08\,\fm$. With the current accuracy, both unitarity
    violations and more standard discretisation effects are found to
    be small for this quantity.}
  \label{fig:charm_etmc}
\end{figure}

A similar mixed action with $\nft$ Wtm sea fermions and OS valence
quarks was employed by ETMC~\cite{Constantinou:2010qv}. The motivation
for such a setup is to determine the Kaon bag parameter $B_K$ with
Wilson-type fermions by simultaneously preserving $\Oa$ improvement
and absence of wrong chirality operator
mixings~\cite{Frezzotti:2004wz,Constantinou:2010qv,Dimopoulos:2010wq}. The
mixed action OS mesons were constructed by using the same sign of
$r_f$, $r_1=r_2$, in the two quark propagators, while the unitary
mesons had opposite signs, $r_1=-r_2$. Fig.~\ref{fig:mk_nf2_etmc}
shows the continuum limit scaling of $m_K^2$ for the unitary and mixed
action mesons~\cite{Constantinou:2010qv}. The scaling behaviour is
consistent with $\Oa$ improvement and unitarity violations in the
mixed action are observed to vanish in the continuum
limit. Significantly larger cut-off effects are reported in
Fig.~\ref{fig:mk_nf2_etmc} when the meson is built of quarks with the
same signs of $r_f$ (red squares). A good scaling behaviour is
observed instead in the case of $f_K$~\cite{Constantinou:2010qv}. For
the meson with valence quarks having $r_1=-r_2$, the scaling of both
$m_K^2$ (blue circles) and $f_K$ show only moderate $\Oasq$ lattice
artifacts~\cite{Constantinou:2010qv}.

When considering the $\nftoo$ case, a mixed action can be employed in
the strange and charm sectors. The mixed action Kaon and $D$-meson include
$s$ and $c$ sea quark effects arising the action in eq.~(\ref{eq:sf})
while OS quarks, with $r_1=-r_2$, are used in the valence
sector. Fig.~\ref{fig:fk_ma_etmc} compares unitary and mixed action
determinations of $f_K$~\cite{Farchioni:2010tb} once the sea and
valence strange quark masses have been matched via the Kaon mass. No
signs of large unitarity violations are observed in this
quantity. Additionally, the values of $f_K$ from two lattice spacings,
$a\approx 0.06$ and $0.08\,\fm$, agree within errors, suggesting that
discretisation effects are small.

An SU(2) $\chi$PT fit together with an interpolation in the valence
sector to the experimental value of $2m_K^2-m_\pi^2$, provides
preliminary estimates of $f_K=160(2)\,\mev$ and
$f_K/f_\pi=1.224(13)$~\cite{Farchioni:2010tb}, where only statistical
errors are quoted.\,\footnote{As in the light sector, the experimental
  value of $f_\pi$ is used to set the scale.}  These $\nftoo$
estimates are compatible with other recent results from $\nft$ and
$\nfto$ calculations~\cite{Hoelbling:2011kk,Colangelo:2010et}. First
determinations of the pseudoscalar meson masses and decay constants in
the charm sector~\cite{Farchioni:2010tb} and of the low-lying baryon
spectrum~\cite{Drach:2010hy} were reported at this conference.

\section{Non-Perturbative Renormalisation}
\label{sec:npr}

Many observables, including fundamental parameters of QCD such as the
strong coupling constant or the quark masses, require a
renormalisation procedure. On the lattice, the removal of UV
divergences can in principle be performed by means of lattice
perturbation theory. However, in practise, the truncation errors can
be large due to the poor convergence of the series and to difficulties
in determining higher order terms. Non-perturbative renormalisation
allows to overcome these problems.

The absence of mass-dependence in the anomalous dimensions entering in
the renormalisation group equations is guaranteed in mass-independent
renormalisation schemes~\cite{Weinberg:1951ss}. In lattice
computations, this implies that the effects of all active quark
flavours need to be incorporated and that the chiral limit of each of
the quark masses has to be taken. When aiming at the renormalisation
of scale-dependent observables determined from $\nfto$ and $\nftoo$
simulations, the chiral limit of the strange and charm quark masses
has to be taken during renormalisation implying that dedicated
simulations with $\nfth$ and $\nff$ degenerate flavours, respectively,
need to be performed. Keeping the quark masses fixed to their physical
values would introduce a systematic effect at the level of
renormalisation. Due to its large mass, the charm quark would tend to
decouple from the dynamics while, in a mass-independent scheme, it
should behave as an active quark.

The non-perturbative renormalisation schemes currently in use are the
Schr\"odinger Functional (SF)~\cite{Luscher:1992an} and the
RI-MOM~\cite{Martinelli:1994ty} schemes. We refer
to~\cite{Aoki:2010yq} for a review on recent developments on these
schemes and variants of them. In the SF it is possible to work
directly with massless quarks while a chiral extrapolation has to be
performed in the case of RI-MOM. Several groups have recently
performed dynamical simulations with $\nfth$ and $\nff$ degenerate
flavours. 

The running of the coupling~\cite{Aoki:2009tf} and of the quark
mass~\cite{Aoki:2010wm} with $\nfth$ flavours has been determined
through the step-scaling function in the SF scheme by PACS-CS using
the Iwasaki gauge action and non-perturbatively $\Oa$ improved Wilson
fermions. Dedicated $\nfth$ simulations have been performed by BMW to
renormalise the quark mass in the RI-MOM scheme~\cite{Durr:2010aw}
with the tree-level improved Symanzik gauge action and tree-level
improved Wilson fermions coupled to gauge links with two levels of HEX
smearing.

\begin{figure}[t!]
  \vspace*{-0.5cm}
  \centering
  \subfigure[\label{fig:rimom_etmc2}]{
    \includegraphics[width=0.485\linewidth]{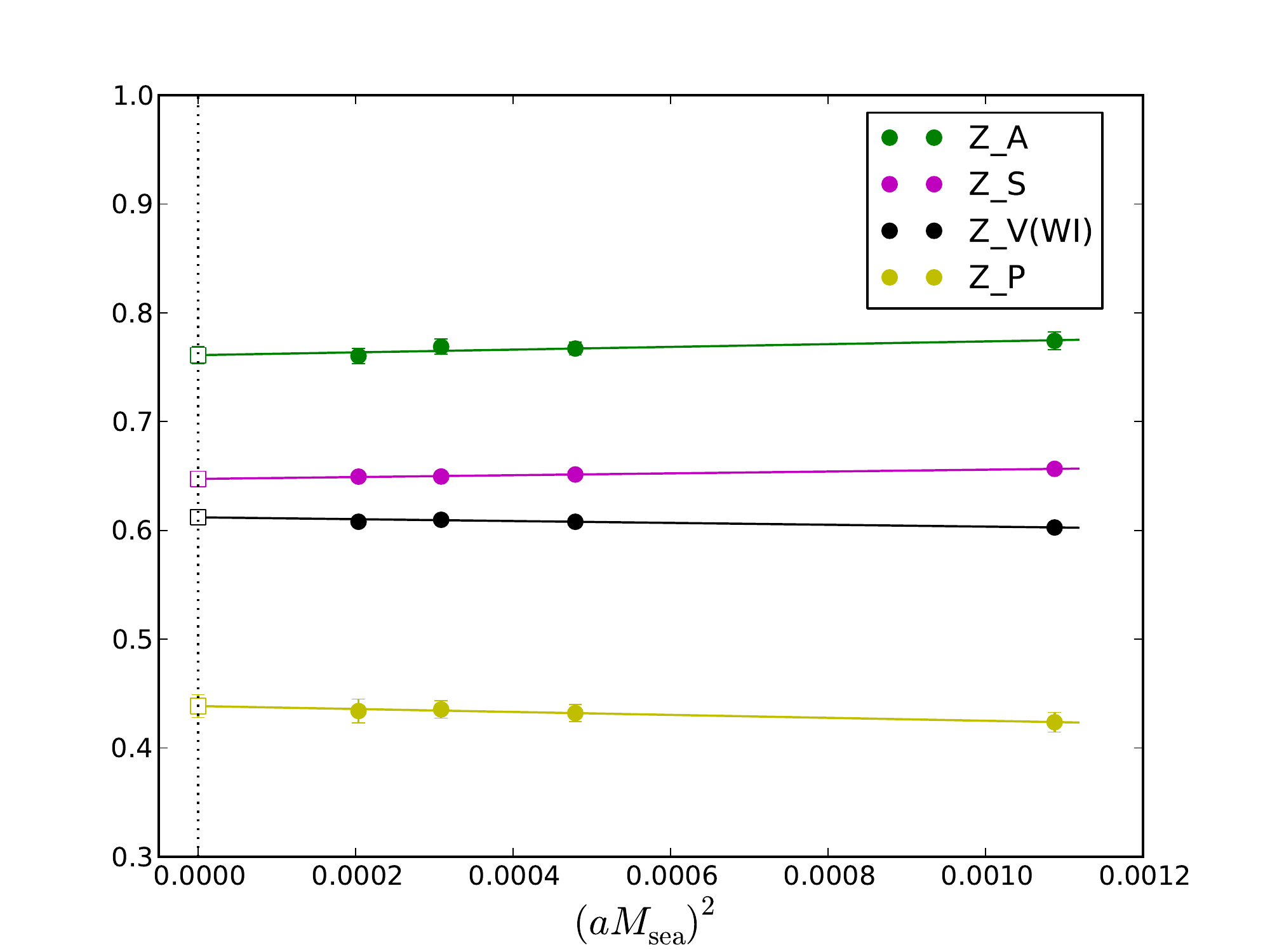}}
  \quad
  \subfigure[\label{fig:coupling_alpha}]{
    \includegraphics[width=0.455\linewidth]{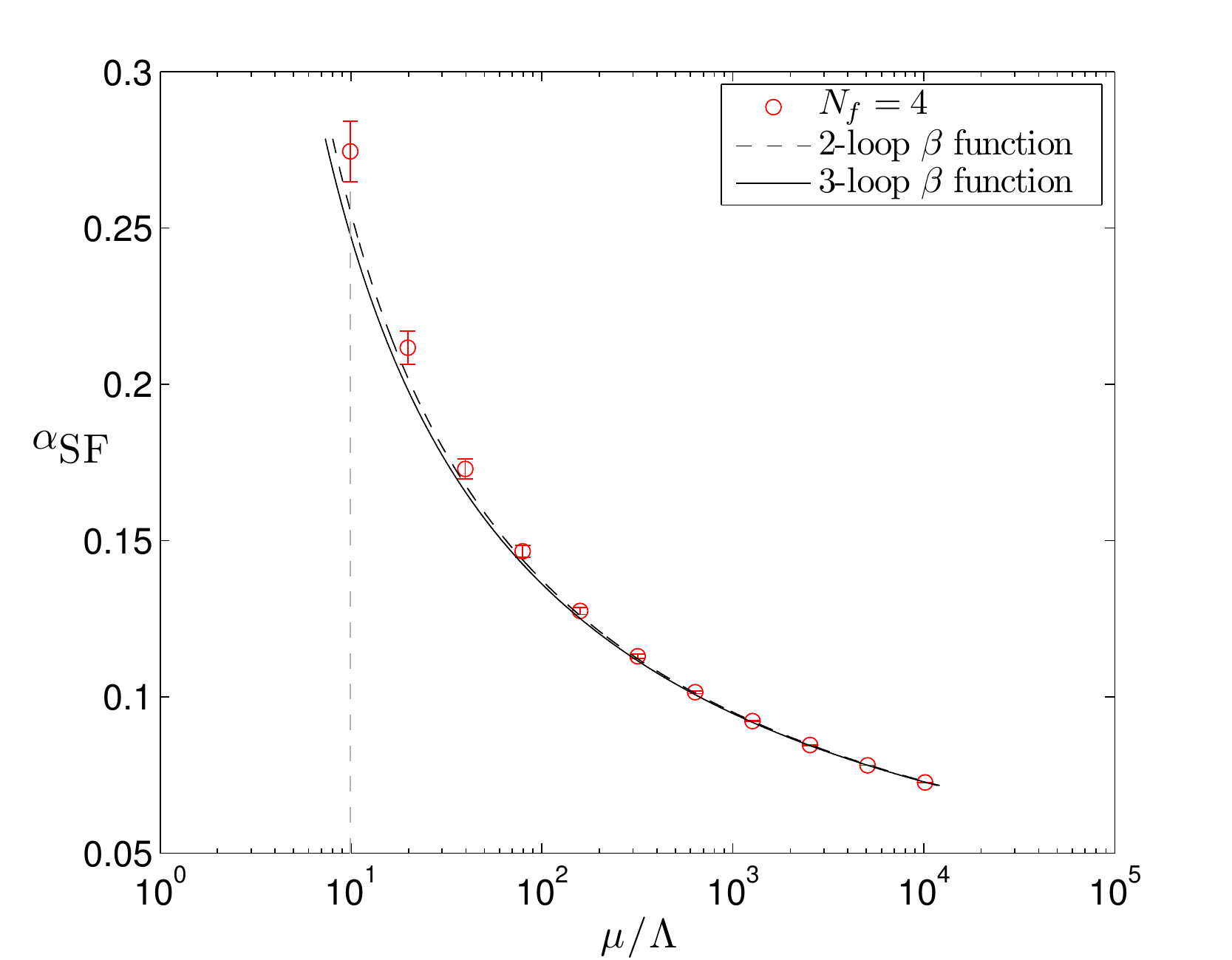}}
  \caption{(a) Sea quark mass dependence of quark bilinear
    renormalisation factors from $\nff$ ETMC calculations in the
    RI-MOM scheme at $a \approx 0.08\,\fm$. (b) Non-perturbative
    running of the $\nff$ QCD coupling in the SF scheme as determined
    by the ALPHA collaboration. A comparison to perturbation theory is
    shown.}
  \label{fig:renorm}
\end{figure}

In the four-flavour case, ETMC is currently generating ensembles with
two degenerate doublets of Wtm fermions\,\footnote{The fermionic
  action for each of the doublets corresponds to eq.~(\ref{eq:sl}).}
and the Iwasaki gauge action to perform the renormalisation in RI-MOM
scheme~\cite{Dimopoulos:2011wz}. In these $\nff$ simulations, the PCAC
quark mass, $m_{\rm PCAC}$, does not show a sufficiently smooth
dependence on the untwisted quark mass $m_{0,l}$ around its critical
value, for $a \gtrsim 0.08\,\fm$ and a twisted mass $\mu_\ell \approx
0.4m_s$. Indeed, for simulations close to maximal twist, i.e. $\mpcac
\approx 0$, the integrated autocorrelation time of $\mpcac$
significantly grows. Simulating at maximal twist would therefore
require fairly long runs to reliably control the value and error of
$\mpcac$. The stability of the simulations improves when working out
of maximal twist, i.e at non vanishing values of both the standard
(untwisted) mass and the twisted mass parameter. In this case, the
renormalised quark mass is given by the polar mass, $\widehat M =
Z_{\rm P}^{-1} \sqrt{Z_{\rm A}^2 \mpcac^2 + \mu_\ell^2}$, which should
eventually be extrapolated to the chiral limit. The $\Oa$ improvement
of the renormalisation factors is achieved by averaging estimators
from simulations with equal $\widehat M$ but opposite values of
$\mpcac/\mu_\ell$~\cite{Dimopoulos:2011wz}. Fig.~\ref{fig:rimom_etmc2}
shows the sea quark mass dependence of quark-bilinear renormalisation
factors from a preliminary analysis at $a \approx 0.08\,\fm$. The mild
dependence on the sea quark mass is in line with ETMC studies in the
two-flavour case~\cite{Constantinou:2010gr}.

The running up to high energy scales of the QCD coupling with $\nff$
flavours has been recently determined non-perturbatively in the SF
scheme by two groups. The study by the ALPHA
collaboration~\cite{Tekin:2010mm,Sommer:2010ui} uses the Wilson
plaquette gauge action and non-perturbatively $\Oa$ improved Wilson
fermions. Given the current statistical precision, cutoff effects in
the step-scaling function are observed to be small at the two finer
lattice spacings. Fig.~\ref{fig:coupling_alpha} shows the
non-perturbative running of the coupling and a comparison to
perturbation theory. The observed deviations from perturbation theory
at low-energies (vertical dashed line), is a warning about the use of
perturbation theory in a regime where the coupling is not sufficiently
weak. The running of the coupling is also being studied by S.~Sint and
P.~P\'erez Rubio~\cite{PerezRubio:2010ke} using the plaquette gauge
action and a single staggered fermion field, corresponding to four
flavours of massless quarks. Two regularisations differing by the time
extent, $T=L \pm a$, are used to estimate systematic effects due to
lattice artifacts. Indeed, in this way, non-negligible cutoff effects
are identified in the step-scaling
function~\cite{PerezRubio:2010ke}. It is interesting to note that
these two studies of the $\nff$ running coupling can be used to check
the universality of the step-scaling function. A natural extension of
these investigations is the determination of the $\Lambda$ parameter,
which requires the input of a physical quantity determined from
$\nftoo$ simulations.

\section*{Conclusions}

Several studies have recently been devoted to lattice QCD simulations
with four dynamical flavours. First physical results from observables
in the light, strange and charm sectors have been presented at this
conference. As an important check of this new simulation setup, the
study of light-quark observables, such as the light pseudoscalar decay
constant or the nucleon mass, revealed a good continuum-limit scaling
behaviour, with no clear evidence of large cutoff effects coming from
the heavy charm quark in the sea. The overall computational cost of
the four flavour simulation program is larger than that of $\nft$ or
$\nfto$ calculations due to further tuning effort of the quark
masses. Furthermore, dedicated simulations for non-perturbative
renormalisation need to be performed. The fact that several lattice
groups have already started these calculations clearly indicates that
such a computational effort is not a strong limitation to proceed with
four-flavour QCD simulations. Studies of the continuum-limit scaling
of observables containing valence charm quarks and of quantities
requiring renormalisation are among the open questions which still
need to be addressed in this novel setup.

\section*{Acknowledgements}

\noindent I thank P.~Dimopoulos, R.~Frezzotti, K.~Jansen and C.~Urbach
for very useful suggestions and discussions and for comments on this
manuscript. I wish to thank P.~Boyle, T.W.~Chiu, N.~Christ,
P.~Dimopoulos, V.~Drach, F.~Farchioni, R.~Frezzotti, E.~Garc\'ia
Ramos, S.~Gottlieb, C.~Hoelbling, K.~Jansen, C.~Lang, V.~Lubicz,
C.~McNeile, C.~Michael, I.~Montvay, D.~Palao, P.~P\'erez Rubio,
D.~Pleiter, S.~Reker, D.~Renner, G.C.~Rossi, F.~Sanfilippo, S.~Sharpe,
A.~Shindler, S.~Sint, R.~Sommer, C.~Tarantino, D.~Toussaint,
C.~Urbach, M.~Wagner, U.~Wenger and all my colleagues from ETMC for
help in preparing this work.

\bibliographystyle{h-physrev5}
\bibliography{bibliography}

\end{document}